\documentclass[aps,10pt,tighten,twocolumn,showpacs]{revtex4}%
\usepackage{amssymb}
\usepackage{graphicx}
\usepackage{amsmath}
\usepackage{amsfonts}%
\providecommand{\U}[1]{\protect\rule{.1in}{.1in}}
\providecommand{\U}[1]{\protect\rule{.1in}{.1in}}
\providecommand{\U}[1]{\protect\rule{.1in}{.1in}}
\providecommand{\U}[1]{\protect\rule{.1in}{.1in}}
\providecommand{\U}[1]{\protect\rule{.1in}{.1in}}
\providecommand{\U}[1]{\protect\rule{.1in}{.1in}}

\begin{document}
\title[ ]{Spin polarization control through resonant states in an Fe/GaAs Schottky barrier}
\author{S. Honda$^{1}$, H. Itoh$^{2}$, and J. Inoue$^{1}$}
\affiliation{$^{1}$Department of Applied Physics, Nagoya University, Nagoya 464-8603, Japan}
\affiliation{$^{2}$Department of Pure and Applied Physics, Kansai University, Suita
564-8680, Japan}
\author{H. Kurebayashi$^{3}$, T. Trypiniotis$^{3}$, C. H. W. Barnes$^{3}$, A.
Hirohata$^{4}$, and J. A. C. Bland$^{3}$}
\affiliation{$^{3}$Cavendish Laboratory, University of Cambridge, J. J. Thomson Ave.,
Cambrigde, CB3 0HE, UK,}
\affiliation{$^{4}$Department of Electronics, University of York, Heslington, York YO10
5DD, UK.}

\begin{abstract}
Spin polarization of the tunnel conductivity has been studied for Fe/GaAs
junctions with Schottky barriers. It is shown that band matching of resonant
interface states within the Schottky barrier defines the sign of spin
polarization of electrons transported through the barrier. The results account
very well for experimental results including the tunneling of photo-excited
electrons, and suggest that the spin polarization (from -100\% to 100\%) is
dependent on the Schottky barrier height. They also suggest that the sign of
the spin polarization can be controlled with a bias voltage.

\end{abstract}

\maketitle

One of the main aims in semiconductor (SC) spintronics is to use the spin
degree of freedom of electrons for novel electronic devices. The use of
ferromagnetic (FM) contacts to inject a spin-polarized current into a SC has
been intensively studied as a means to achieve spintronic control in SC
devices, and has led to many successful experiments that demonstrate a
spin-polarized current through the contact
\cite{Motsnyi02,Hanbicki03,Andresen03,Erve04,Adelmann05,Jiang05,Steinmuller05}%
. The spin-injection efficiencies measured in these experiments are impressive
with the highest being 57\% at 100K \cite{Jiang05}, but they are not as high
as first-principles band calculations predict \cite{Wunnicke02}. Quite
recently, negative spin polarization (negative $P$) of the tunnel current
through the Schottky barrier of FM/GaAs has been reported in several
experiments; observation of spin accumulation in lateral Fe/GaAs/Fe
\cite{Cooker05,Lou07}, imaging of injected spins in FeCo/GaAs junctions
\cite{Kotissek07}, measurements of tunnel magnetoresistance in Fe/GaAs/Fe
junctions \cite{Moser06}, and spin-filtering experiments with photo-excited
electrons produced in the GaAs layer \cite{Kurebayashi07}. The bias dependence
of negative $P$ in these experiments, however, is still controversial as
argued by Lou et al. \cite{Lou07}, suggesting that the band structure at the
FM/SC interface and the Schottky barrier may play a key role in determining
spin transport across the interface.

Several mechanisms of negative $P$ have been proposed for tunneling
conductance of Fe junctions; resonant tunneling via extrinsic impurity levels
in the barrier region \cite{Tsymbal03}, and interface resonant states (IRSs)
appeared intrinsically in the minority spin state of Fe \cite{Tiusan04}. The
extrinsic mechanism may be ruled out for Fe/GaAs junctions since the negative
$P$ appears in ideal interfaces \cite{Moser06}. As for IRSs mechanism, Chanits
\textit{et al}. have proposed that IRSs at Fe/GaAs interfaces are responsible
for the negative $P$ by performing a first-principles calculation for an
Fe/GaAs/Cu junction without the Schottky barrier \cite{Chantis07}. On the
other hand, Dery and Sham proposed that the sign of $P$ is governed by a
competition between conduction electron tunneling with positive $P$ and
tunneling of localized electrons with negative $P$ in an over-doped layer near
the Fe/GaAs interface \cite{Dery07}. In addition, Dery-Sham have proposed a
novel device of spin-switch by a gate voltage control. In spite of these
important works, one should examine the spin transport mechanism further,
since the role of the Schottky barrier is still unclear and the negative $P$
has been observed also in Fe/GaAs junctions without the over-doped layer
\cite{Moser06,Kurebayashi07}.

In this Letter, we will show that \textit{the IRSs within the Schottky
barrier} play an important role for the negative $P$ and its bias dependence.
Because of the band symmetry of both Fe and GaAs layers, and symmetry
dependent hybridization with the spin-polarized Fe states, down ($\downarrow$)
spin IRSs appear near the bottom of GaAs conduction band and vary the
thickness of the Schottky barrier effectively. Due to the features of IRSs in
the Schottky barreir, a sharp variation of $P$ from $\sim100$\% to $\sim
-100$\% occurs when the energy of incident electrons or the applied bias is
changed. The present results not only explain the spin-filtering of
photo-excited electrons in the GaAs layer semi-quantitatively, but may resolve
the controversy about the bias dependence of negative $P$ observed. A strong
variation of $P$ in a small bias region may also suggest possible control of
spin polarization with a bias voltage, or with Schottky barrier height using
FM alloys with different work functions. The former phenomenon could be used
to make a new type of spin-switch devices.

In the following, we calculate the spin-dependent tunnel conductance for
photo-excited electrons through an Fe/GaAs contact with a Schottky barrier
using a full-orbital tight-binding model and the linear response theory. The
present model is sufficiently realistic to reproduce the previous results
obtained in the first principles \cite{Wunnicke02}, and feasible to deal with
the thick tunnel barrier formed by a Schottky barrier. The model is also
appropriate to study the effects of the IRSs on the tunnel conductance
\cite{Honda06}.

When we restrict our discussion to absolute zero temperature and neglect a
thermally excited Schottky current, the tunnel current $I_{L(R)}$ of electrons
excited by left(right) circularly polarized light in GaAs can easily be
obtained by using the selection rule, the symmetry of the valence and
conduction bands \cite{Hirohata07}, and spin ($\sigma=\uparrow,\downarrow$)
dependent tunnel conductance $\Gamma_{\sigma}(E)$ at an energy $E$. When the
Schottky barrier is sufficiently high and thick, the tunnel currents should be
governed by the tunneling probability at the excitation energy that
$E_{1}=E_{ph}-E_{g}$ and $E_{0}=E_{ph}-E_{g}-\Delta$, from $P_{3/2}$ and
split-off $P_{1/2}$ valence bands, respectively, where $E_{ph}$, $E_{g}$ and
$\Delta$ are the photon energy, the band gap energy and the spin splitting of
the valence band, respectively. Then, the difference between $I_{L}$ and
$I_{R}$ under a forward bias $V_{F}$ is given by
\begin{equation}
\Delta I\equiv I_{L}-I_{R}\sim2\left[  P(E_{1})\Gamma(E_{1})-P(E_{0}%
)\Gamma(E_{0})\right]  V_{F},
\end{equation}
where $P(E)=\left[  \Gamma_{\uparrow}(E)-\Gamma_{\downarrow}(E)\right]
/\left[  \Gamma_{\uparrow}(E)+\Gamma_{\downarrow}(E)\right]  $ is the spin
polarization of the tunneling conductance.

Three different photon energies were used to excite the valence electrons,
$E_{ph}=1.58,1.85$ and 1.96eV, which give $E_{1}=0.15$ $(\equiv\varepsilon
_{1}),0.42$ $(\equiv\varepsilon_{2})$, and 0.53eV$(\equiv\varepsilon_{3})$,
respectively. Since the values of $E_{0}$ are smaller than $E_{1}$ by
$\Delta=0.35$eV, we expect $\Gamma(E_{1})\gg\Gamma(E_{0})$ unless the Schottky
barrier is too low. The experimental results of the differential tunnel
conductance $\Delta I/V_{F}$ show that the sign of $\Delta I/V_{F}$ for
$E_{ph}=1.58$eV is different from that for $E_{ph}=1.85$ and 1.96eV, and that
$|\Delta I/V_{F}|$ begins to decrease when $V_{F}$ exceeds 0.2V
\cite{Kurebayashi07}.

The tunnel conductance has been calculated by using a full-orbital
tight-binding model; $s$, $p$ and $d$ orbitals for Fe, and $s$ and $p$
orbitals for GaAs. The hopping parameters are determined by fitting the
calculated energy dispersion curves to those obtained by the other
calculations \cite{Papa86,Talwar82}. The local density of states (DOS) at each
layer and the tunnel conductance at an energy $E$ are calculated by using
recursive Green's function method. We calculate the tunnel conductance for an
Fe/GaAs(001) interface with both Fe-As and Fe-Ga contacts, neglecting the
mismatch of the lattice constants between Fe and GaAs.

\begin{figure}[tbh]
\begin{center}
\includegraphics[width=0.9\linewidth]{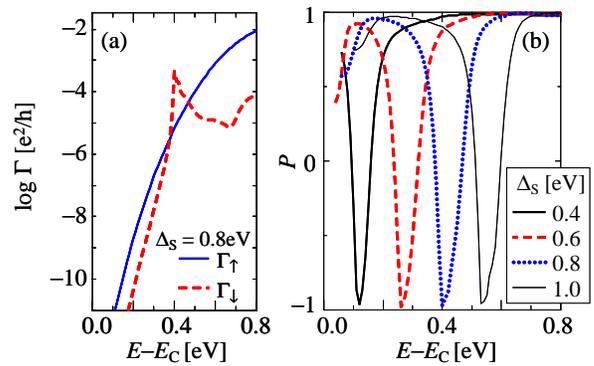}
\end{center}
\caption{(color online) (a) Calculated results of conductance as a function of
an energy for the Fe-As contact with $\Delta_{S}=0.8$, $L_{S}=400$ML and zero
bias, and (b) those of the spin polarization for various values of $\Delta
_{S}$. }%
\end{figure}

We adopt a model in which the shape of the Schottky barrier (the position
dependence of the bottom of the conduction band) is given by $E_{C}%
(\ell)=\Delta_{S}e^{-\ell/\lambda}$, where $\Delta_{S}$ and $\ell$ are the
Schottky barrier height and the distance measured from the interface. The
value of $\lambda$ is determined in such a way that $E_{C}(\ell)$ becomes
10$^{-4}$eV at $\ell=L_{S}$. The Fermi level $E_{F}$ of bulk GaAs is taken to
be the bottom of the conduction band $E_{F}=E_{C}(\ell>L_{S})$ assumed for
highly doped n-type GaAs. The forward bias dependence is taken into account by
shifting the GaAs bands by $eV_{F}$, i.e, $E_{C}(\ell)\longrightarrow
(\Delta_{S}-eV_{F})e^{-\ell/\lambda}+eV_{F}$. Bias dependence of the barrier
thickness is neglected since its effect is much smaller than that of the
reduction of the effective barrier height. In the practical calculations, the
Schottky barrier is included as a position-dependent shift of the atomic
potential of Ga and As atoms. Calculated results of the tunnel conductance and
spin polarization for incident electrons normal to the layer plane agree
semi-quantitatively with those obtained in the first principles
\cite{Wunnicke02}.

Figure 1(a) shows the calculated results of the spin-resolved conductance
$\Gamma_{\sigma}(\sigma=\uparrow,\downarrow)$ for an Fe-As contact with a
Schottky barrier with $L_{S}=400$ML and $\Delta_{S}=0.8$eV. We see that
$\Gamma_{\uparrow}$ increases nearly monotonically, while $\Gamma_{\downarrow
}$ shows a sharp peak around $E-E_{C}=0.4$eV. Therefore, the spin polarization
of the tunnel conductance becomes negative in a specific energy window.
$\Gamma_{\downarrow}$ is nearly constant for $E-E_{C}\gtrsim0.8$eV (not shown)
until the energy $E$ touches the $\Delta_{1}$ band in the Fe minority spin
states. When $E-E_{F}\sim1$eV, $\Gamma_{\downarrow}$ increases rapidly as the
$\Delta_{1}$ band of the Fe minority spin state begins to contribute the
tunneling, resulting in an abrupt decrease of the spin polarization at the
energy. Figure 1(b) shows the spin polarization of the tunnel conductance for
various values of $\Delta_{S}$. We find that $P$ can be $\sim-100$\% for a
certain energy window, and that the energy window shifts in proportion to
$\Delta_{S}$. It should be noted that the calculated conductance is less
accurate when $E-E_{C}\sim0$, since the thickness of the Schottky barrier at
this energy region is too thick for numerical calculations.

\begin{figure}[tbh]
\begin{center}
\includegraphics[width=0.8\linewidth]{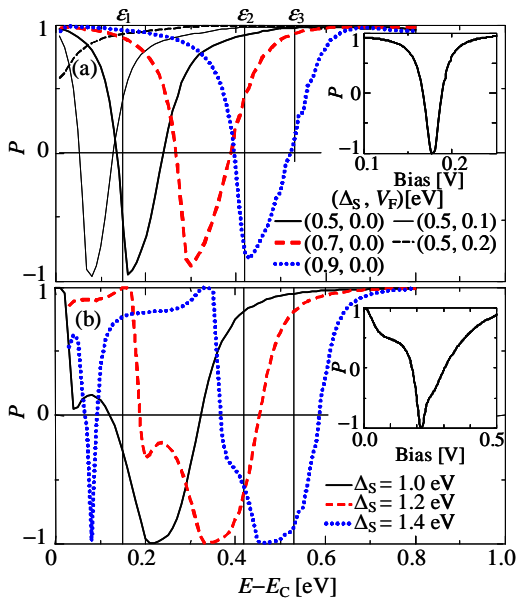}
\end{center}
\caption{(color online) Calculated results of the spin polarization of the
tunnel conductance as a function of an energy for (a) the Fe-As contact with
$L_{S}=200$ML and vairous values of $\Delta_{S}$ at a bias voltage $V_{F}$,
and for (b) the Fe-Ga contact with $L_{S}=200$ML and various values of
$\Delta_{S}$ at zero bias. Insets of (a) and (b) are the bias dependence of
$P$ at $E=E_{C}+0.005$eV for the Fe-As contact with $\Delta_{S}=0.5$eV and the
Fe-Ga contact with $\Delta_{S}=1.0$eV.}%
\end{figure}

Calculated results of $P$ for the Fe-As and Fe-Ga contacts with $L_{S}=$200ML
are shown in Figs. 2(a) and 2(b), respectively. In Fig. 2(a) the negative spin
polarization becomes less perfect when $L_{S}=$200ML. This is because $L_{S}$
is small, and more states in the Fermi surface begin to contribute to the
tunneling. Similar to the results for $L_{S}=400$ML, the peaks of the negative
$P$ shift to the lower energy region with decreasing $\Delta_{S}$. Thin curves
in Fig. 2(a) show the bias dependence of the spin polarization for the Fe-As
contact with $L_{S}=200$ML and $\Delta_{S}=0.5$eV. We see the energy window
with negative $P$ shifts towards the lower energy region in proportion to the
bias voltage $V_{F}$.

The energy dependence of $P$ for the Fe-Ga contact is essentially the same
with that for the Fe-As contacts, however, there are a few differences to be
noted: (i) The energy windows for the negative $P$ are wider for the the Fe-Ga
contacts than those for the Fe-As contacts. (ii) The negative spin
polarization for the Fe-Ga contacts can always be perfect irrespective to
$L_{S}$. (iii) Most importantly, a large value of $\Delta_{S}$ is necessary to
realize the negative $P$ for the Fe-Ga contacts.

\begin{figure}[tbh]
\begin{center}
\includegraphics[width=1.0\linewidth]{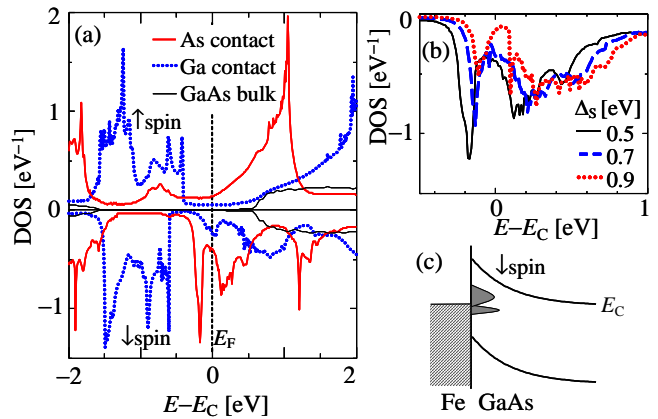}
\end{center}
\caption{(color online) (a) Calculated results of the local DOS of As and Ga
at the interface for the Fe-As and Fe-Ga contacts, respectively, with
$\Delta_{S}=0.5$eV, $L_{S}=200$ML and zero bias. (b) Enlargemnet of the
$\downarrow$ spin local DOS for the Fe-As contact, and (c) a schematic figure
of the resonant states in the Schottky barrier. }%
\end{figure}

The above mentioned results can well be accounted for in terms of the IRSs in
the Schottky barrier of the GaAs layer. Figure 3(a) presents the local DOS on
the As and Ga layers at the Fe-As and Fe-Ga contacts, respectively, with the
Schottky barrier of $L_{S}=200$ML and $\Delta_{S}=0.5$eV. We find many sharp
peaks appear in both the As and Ga local DOS, which may be identified to be
the IRSs. These IRSs are spin dependent due to the hybridization with the
spin-polarized Fe bands. The existence of an IRS at $E-E_{C}\sim0.2$eV in the
$\downarrow$ spin state may explain the negative value of $P$ calculated for
the Fe-As contact with these parameter values. As $\Delta_{S}$ increases, the
IRS is shifted by nearly the same amount of the increase of $\Delta_{S}$ as
shown in Fig. 3(b). These results are in good accordance with the shift of the
energy window where $P<0$. A schematic figure of the IRSs in the present model
is shown in Fig. 3(c). When a forward bias $V_{F}$ is applied, the chemical
potential of the GaAs layer (in other words, $E_{C}$) shifts by $eV_{F}$, and
therefore the energy window of negative $P$ is shifted to the lower energy
region by $\sim eV_{F}$ as shown in Fig. 2(a).

Since the IRSs are formed by an interference effect between the incident and
reflected waves of the conduction band of GaAs at the interface, they are
dominated by the $\Delta_{1}$ symmetry for the Fe/GaAs(001) interface.
Therefore, they hybridize stronger with $\uparrow$ spin Fe bands which have
the $\Delta_{1}$ symmetry band near $E_{C}$ than with $\downarrow$ spin Fe
bands. Strong hybridization in the $\uparrow$ spin states pushes down (up) the
bonding (anti-bonding) state of the IRSs, resulting in a weak intensity of the
IRSs near $E_{C}$. The IRSs in the $\downarrow$ spin state with $k_{\Vert}%
\neq(0,0)$, where $k_{\Vert}$ is a momentum parallel to the layer plane,
hybridize with the $\Sigma_{1}$ band of Fe mainly, and have rather strong
intensity near $E_{C}$ as shown in Fig. 3(a). Although the IRSs are evanesent
states, they make the effective barrier thickness thinner significantly,
therefore giving rise to the negative $P$. It should also be noted the nature
of the IRSs is changed with different layer stacking orientation, since the
IRSs are symmetry dependent.

\begin{figure}[tbh]
\begin{center}
\includegraphics[width=1.0\linewidth]{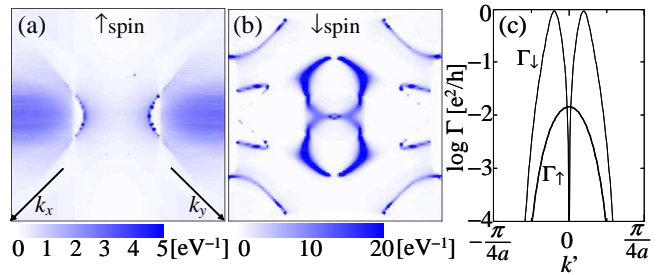}
\end{center}
\caption{(color online) Momentum-resolved density of states for the (a)
$\uparrow$ and (b) $\downarrow$ spin states at $E-E_{C}=0.2$eV, and (c)
momentum-resolved conductance for the $\uparrow$ and $\downarrow$ spin states
for the Fe-As contact, where $k^{\prime}$ indicates the momentum along
$k_{x}-k_{y}$. Parameter values are $L_{S}=200$ML and $\Delta_{S}=0.5$eV. }%
\end{figure}

Above consideration is justified by the calculated results of the $k_{\Vert}%
$-resolved local DOS and conductance, which are shown in Fig. 4. Figures 4(a)
and (b) are the local DOS of the $\uparrow$ and $\downarrow$ spin states of
the As layer at the Fe-As contact. The both local DOS spread over the whole
Brillouin zone, however, the $\downarrow$ spin local DOS is much larger than
the $\uparrow$ spin one near $k_{\Vert}=(0,0)$. Since the Schottky barrier is
thick, the tunnel conductance is governed by the states near $k_{\Vert}%
=(0,0)$, and as a result $\Gamma_{\downarrow}$ becomes much larger than
$\Gamma_{\uparrow}$ as shown in Fig. 4(c). It should be noted that the
$\Gamma_{\downarrow}$ is precisely zero at $k_{\Vert}=(0,0)$ by symmetry.

It should also be noted that all of the resonant states in the Fe minority
spin state do not contribute to the tunneling via the IRSs formed at an
Fe/GaAs(001) interface, since the former states may have $\Delta_{5}$
symmetry, among which only $p_{x}$ and $p_{y}$ orbitals hybridize with the
$\Sigma_{1}$ band when $k_{\Vert}\neq(0,0)$. We have confirmed that the
resonant states in the Fe minority spin state (not shown) stay at almost the
same energy position when $\Delta_{S}$ is increased. Since it is difficult to
explain the shift of the energy window proportional to $\Delta_{S}$ by the
resonant states in the Fe layer, it would less contribute to the origin of the
negative spin polarization calculated here.

Now let us compare the calculated results with experimental ones. As
mentioned, the experiments have used three excitation energies $\varepsilon
_{1}$, $\varepsilon_{2}$ and $\varepsilon_{3}$, which are shown by vertical
lines in Figs. 2(a) and (b). The experimental results suggest that the sign of
the differential conductance $\Delta I/V_{F}=(I_{L}-I_{R})/V_{F}$ at
$E=\varepsilon_{1}$ is different from that at $E=\varepsilon_{2}$ and
$\varepsilon_{3}$. One of the conditions which agree with the experimental
observation is $\Delta_{S}\sim0.5$eV for the Fe-As contact irrespective to the
barrier thickness, where $P(\varepsilon_{1})<0,P(\varepsilon_{2})>0$ and
$P(\varepsilon_{3})>0$. When the bias voltage is increased to 0.2V,
$P(\varepsilon_{1})$ calculated changes the sign, and $P(\varepsilon_{1})$,
$P(\varepsilon_{2})$ and $P(\varepsilon_{3})$ are all close to 1. The latter
result may not agree with the experimental one in which $|\Delta I/V_{F}|$
begins to decrease above $V_{F}\sim0.2$V. The discrepancy may be attributed to
the quality of the Schottky barrier of the measured sample. The height of the
Schottky barrier estimated experimentally for our sample is 0.23eV, and hence
the electron conduction becomes metallic-like when $V_{F}>0.2$V, leading to a
decrease of the spin polarization across the interface. In addition, the
estimated barrier height 0.23eV can be the lower limit, assuming an in-plane
barrier height distribution where lower barrier (less resistive) parts would
predominate the electron transport property. Consequently, higher barrier
regions in our junction would still give rise to spin-polarized tunneling,
though its weight may decrease. Actually we observed no sign change in $\Delta
I$ for a sample with\ the lower barrier height of 0.1eV. We expect that the
negative spin polarization of the spin-filtering effect should be clearly seen
for high-quality samples with the higher barrier as estimated to be 0.46eV by
Hanbicki \textit{et al}.\cite{Hanbicki03}.

Since the value of $P$ is strongly dependent on the energy, bias voltage as
well as Schottky barrier height as shown in Figs. 2(a), (b) and the insets,
the results could shed light on the enigmatic topic on $P$ at Fe/GaAs
interfaces mentioned in the introduction, and propose a feasible control of
$P$ at an FM/GaAs junction. The inevitable variation of Schottky barrier
heights in experimental samples may explain the observed differences in bias
dependence of $P$ in these cases. Since the value of $P$ varies from $-100$\%
to $+100$\% with the bias voltage, the complete spin polarization tuning by
the bias voltage can be realized in ideal FM/GaAs interfaces. Such devices
should be promising since they require no over-doped layers nor complex
structures with gate terminals for switching $P$ \cite{Dery07}. The proposed
spin-switch devices can operate in low bias voltage regions due to the
switchings seen in the insets of Fig. 2. Control of the interface spin
polarization with different Schottky barrier heights may also be possible by
using FM alloys with different work functions as performed in FM/Si interfaces
\cite{Min08}.


In conclusion, we have calculated the spin polarization of the tunnel
conductivity using a full-orbital tight-binding model, and have shown that the
interface resonant states within the Schottky barrier in the GaAs layer
influence significantly the spin-dependent tunneling across the interface. It
has been clearly shown that the band matching of the IRSs plays a crucial role
on the spin polarization. The theoretical results account well for earlier
experimental results including the tunneling of photo-excited electrons. The
present results suggest that the spin polarization can be controlled by the
Schottky barrier heights, and that a spin-switch device with bias control may
also be promising. Quantitative performance of the device, however, needs more
quantitative calculations including effects of atomic disorder for example.
\cite{Zwierzycki03,Itoh06}.

The work was supported by the Next Generation Super Computing Project,
Nanoscience Program, MEXT, Japan, Grants-in-Aids for Scientific Research in
the priority area "Spin current" from MEXT, Japan, and Grant-in-Aid for the
21st Century COE "Frontiers of Computational Science".

\end{document}